\documentclass[12pt]{article}
\usepackage{graphicx}
\usepackage{amsmath}
\usepackage{amsfonts}
\usepackage{wasysym}
\usepackage{color}

\normalfont\rmfamily\normalsize\large
\begin{document}

\title{Quasi-classical Langmuir wave collapse\\
in a magnetic field \cite{KuznetsovTuritsyn}}
\author{E.A. Kuznetsov, S.K. Turitsyn}
\date{{\small Institute of Automation and Electrometry, Siberian Branch }\\
{\small the USSR Academy of Sciences, 630090 Novosibirsk, SU }}
\maketitle

\begin{abstract}
The anisotropy due to a magnetic field is shown to result in significant changes in Langmuir collapse. 
Using a variational approach, the quasi-classical collapse phenomenon is investigated analytically.
 A hierarchy of quasi-classical collapses is determined, 
along with the structure of a field in the proximity of a stationary singularity that 
is sustained by the continuous absorption of wave energy from a wave packet.

 \end{abstract}


\section{Introduction}
The hierarchy of wave collapses was first demonstrated in the three-dimensional (3D) nonlinear Schrödinger (NLS) equation 
in 1985 \cite{ZakharovKuznetsovMusher1985, ZakharovKuznetsov1986}. 
This hierarchy combines regimes of compression of regions with higher wave intensities, 
starting from a weak collapse where the energy captured into the singularity is small
 (formally tending to zero), up to a strong collapse where the wave energy absorbed 
in the singularity is finite. It was clarified that the strong collapse regime 
corresponds to quasi-classical compression of the wave packet as a whole.
 In contrast, for the weak collapse, described by a self-similar solution of the 3D NLS,
 the wave field compression is not close to quasiclassical in any sense. 
This circumstance, as shown in \cite{ZakharovKuznetsov1986}, 
is a reason for the instability of strong collapse relative to short-wave perturbations
 resulting in singularities of the weak type developing on the background of quasi-classical compression.

Subsequent works, first \cite{VlasovPiskunovaTalanov}, and then \cite{Malkin, ZakharovKosmatovShvets}, 
showed that after singularity formation, wave capture into the singularity continues.
 This process is similar to the funnel effect or the black hole regime. 
Such a post-collapse becomes more intensive with increasing dimensionality $d$: at $d>4$, 
the process transforming to a quasiclassical stage at dimensions greater than four, 
as explained in \cite{ZakharovKosmatovShvets}. The reason for this is well known: 
it is connected with the increased role of nonlinear effects with increasing dimensionality. 
It is less well known that even weak anisotropy can lead to an increase in nonlinearity.
 Such situation occurs for Langmuir waves in a plasma with sufficiently weak magnetic field ($%
\omega_{ce}\ll \omega_{pe}$) when all changes in the dispersion law $%
\omega=\omega_k$ is expressed as an addition to $\omega=\omega_{pe}$, 
\begin{equation*}
\omega_k=\omega_{pe}\left( \frac 12 k^2 r_d^2 +\frac 12 \frac {\omega_{ce}^2%
}{\omega_{pe}^2}\frac{k^2_{\perp}}{k^2}\right),
\end{equation*}
where $\omega_{ce}$, $\omega_{pe}$ are electron cyclotron and plasma
frequencies respectively, $r_d=V_{Te}/\omega_{pe}$ is Debye radius and $%
k_{\perp}$ is the transverse relative to the magnetic field component of the
wave vector $\mathbf{k}$.

It is seen that the magnetic field is essential only for $%
(kr_{d})^{2}<(\omega _{ce}/\omega _{pe})^{2}$. If one considers weak turbulent 
mechanisms of energy transfer, in which induced scattering on ions is the dominant process, 
then for waves in the region where $kr_{d}<\omega_{ce}/\omega_{pe}$, the distribution 
will first narrow over the angle with a decrease in $k_{\perp}$, and only then will $k_{z}$ decrease. 
This means that as a result, a condensate in the region of $\omega_{k}\rightarrow \omega_{pe}$
 will have a characteristic transverse scale much larger than the 
longitudinal one: $k_{\perp}/k_{z}\ll 1$. Therefore, to describe the modulation instability 
of the condensate and its nonlinear stage, i.e. collapse, it is possible to use the equation
 \cite{KuznetsovSkoric, KrasnoselskikhSotnikov} 
\begin{equation}
\frac{\partial ^{2}}{\partial z^{2}}\left( i\psi _{t}+\psi _{zz}\right)
-\Delta _{\perp }\psi +\frac{\partial }{\partial z}\left( |\psi
_{z}|^{2}\psi _{z}\right) =0  \label{main}
\end{equation}%
The equation for the amplitude $\psi$ of the high-frequency (HF) potential, 
written in dimensionless variables, can be obtained by averaging with respect to the 
rapid time $\omega_{pe}^{-1}$. For the derivation of this equation,
 it is assumed that the low-frequency density fluctuations follow the HF pressure, and $k_{\perp}<<k_{z}$. 
Equation (\ref{main}) belongs to the Hamiltonian class, and it can be written in the following form:
\begin{equation*}
i\frac{\partial ^{2}}{\partial z^{2}}\psi _{t}=-\frac{\delta H}{\delta \psi
^{\ast }},
\end{equation*}%
with the Hamiltonian%
\begin{equation*}
H=\int \left( |\psi _{zz}|^{2}+|\nabla _{\perp }\psi |^{2}-\frac{1}{2}|\psi
_{z}|^{4}\right) d\mathbf{r\equiv }I_{1}+I_{2}-I_{3}.
\end{equation*}%
When transverse dispersion is absent, the equation transforms into the one-dimensional nonlinear 
Schrödinger equation, where all essential dynamics are reduced to the interaction of solitons. 
In the three-dimensional model (\ref{main}), the role of nonlinear effects becomes more significant, 
making collapse possible \cite{KrasnoselskikhSotnikov}. As shown in \cite{KuznetsovSkoric}, 
the enhancement of nonlinear effects manifests itself, particularly in the absence of
 three-dimensional solitons in (\ref{main}). This conclusion
 follows from analyzing the dependence of the Hamiltonian $H$ on the similarity parameters $a$ and $b$
\begin{equation}
H(a,b)=I_{1}/a^{2}+I_{2}a^{2}/b^{2}-I_{3}/(ab^{2}),  \label{scaling-H}
\end{equation}%
under scaling transformations $\psi (z,\mathbf{r}_{\mathbf{\perp}})\rightarrow a^{1/2}b^{-1}\psi (z/a,\mathbf{r}_{\mathbf{\perp}}/b)$ 
that preserve the total energy $N=\int |\psi_{z}|^{2}d\mathbf{r}$. The surface $H=H(a,b)$ in this case 
has no stationary points corresponding to three-dimensional solitons,
 which are solutions of the variational problem $\delta (H+\lambda^{2}N)=0$.
 Another important feature of $H(a,b)$ is its unboundedness from below as 
$a,b\rightarrow 0$. If we assume that longitudinal dispersion ($\sim a^{-2}$) is comparable
 to transverse dispersion ($\sim a^{2}/b^{2}$), i.e., $b=Ca^{2}$ with $C$ being a constant, then 
\begin{equation}
H(a,b)=a^{-2}\left( I_{1}+I_{2}/C ^{2}\right) -a^{-5}I_{3}/C ^{2}.
\label{H-C}
\end{equation}%
It is apparent that the Hamiltonian's unboundedness from below results from the nonlinear term 
(corresponding to $I_{3}$). This unboundedness is one of the criteria for wave collapse existence. 
The wave collapse can be likened to a particle falling in an unbounded self-consistent potential. 
However, this analogy is not entirely accurate since there is a fundamental difference between 
particles and waves. Waves, unlike particles, can be radiated. In the three-dimensional NLS model, 
radiation surprisingly promotes collapse, as demonstrated in \cite{ZakharovKuznetsovMusher1985}.
Let $\Omega (t)$ be a region with a negative Hamiltonian,%
\begin{equation}
H_{\Omega }=\int_{\Omega }\left( |\psi _{zz}|^{2}+|\nabla _{\perp }\psi
|^{2}-\frac{1}{2}|\psi _{z}|^{4}\right) d\mathbf{r<0,}  \label{omega}
\end{equation}%
(we will refer to this region as a cavity in what follows). Then, due to
radiation, the maximum value $|\psi _{z}|^{2}$ can only increase. From (\ref%
{omega}) one can obtain an estimate \cite{KuznetsovSkoric}%
\begin{equation}
\max_{\Omega }|\psi _{z}|^{2}\geq H_{\Omega }/2N_{\Omega }  \label{estimate}
\end{equation}%
Due to radiation from the cavity, $H_{\Omega}$ can only decrease, i.e., 
it becomes even more negative. In this case, $N_{\Omega}$,
 as a positive quantity, tends to zero. As a result, the ratio in (\ref{estimate}) 
tends to infinity. Thus, we can conclude that radiation promotes singularity formation. 
In a sense, this process is similar to the evaporation of a droplet when the droplet 
radius passes through a critical size.
Another feature we want to draw attention to concerns the behavior of the NLS Hamiltonian 
under scaling transformations for different dimensions $d$. Similar to (\ref{scaling-H}), we can obtain for the NLS that
\begin{equation*}
H_{NLS}=\frac{I_{1}}{a^{2}}-\frac{I_{2}}{a^{d}}
\end{equation*}%
where 
\begin{equation*}
I_{1}=\int |\nabla \psi |^{2}d\mathbf{r},\,\,\,I_{2}=\frac{1}{2}\int |\psi
|^{4}d\mathbf{r}
\end{equation*}%
and $a$ is a similarity parameter of the scaling transformations preserving $%
N=\int |\psi _{z}|^{2}d\mathbf{r}$.
By comparing this expression with (\ref{scaling-H}), we can conclude that model (\ref{main}) 
is highly nonlinear, corresponding to the NLSE with dimension $d=5$. 
This comparison serves as an indication that all collapses, except the weakest 
one studied in \cite{KuznetsovSkoric}, and the post-collapse can be treated quasi-classically.

This work focuses on investigating the quasi-classical collapse of Langmuir waves in 
a magnetic field described by equation (\ref{main}). The article is organized as follows. 
In Section 2, we apply the variational method to describe 
strong collapse. Next, we derive quasi-classical equations, determine the hierarchy of 
self-similar collapsing solutions, and investigate their stability. 
In the last section, we present the analytical solution for the quasi-stationary singularity.

\section{Variational approach}

The main characteristic features of the quasi-classical solutions
corresponding to strong collapse can be illustrated using the variational
method. The main idea of an approximate description of the
solutions of equation (\ref{main}) in the framework of this approach is as
follows.

Equation (\ref{main}) can be written by means of the variational principle%
\begin{equation*}
\delta S=0,
\end{equation*}%
where the action $S$ has the form%
\begin{equation}
S=\int \left[ \frac{i}{2}\left( \psi _{z}\psi _{zt}^{\ast }-\psi _{z}^{\ast
}\psi _{zt}\right) +|\psi _{zz}|^{2}+|\nabla _{\perp }\psi |^{2}-\frac{1}{2}%
|\psi _{z}|^{4}\right] dtd\mathbf{r.}  \label{action}
\end{equation}%
By substituting various trial functions into this expression, we can obtain an 
approximate description of the solutions to equation (\ref{main}). 
The accuracy of this approach depends on how well we choose the trial functions.
Let us  substitute into
expression (\ref{action}) a trial function of the following form:
\begin{equation}
\psi (\mathbf{r,}t)=\frac{a}{c}\psi _{0}(z/a,\mathbf{r}_{\perp }/b)\exp
[iz^{2}/2b],  \label{psi}
\end{equation}%
where we assume that parameters $a,b,c$ are functions of $t,$ 
$\psi_0$ is a real function of its arguments that vanishes sufficiently
 rapidly at infinity. The trial function $\psi(\mathbf{r},t)$
 has temporal and spatial oscillations defined by the phase $\Phi=z^2/2b$. 
Its characteristic value is approximately $~a^2/b$. As we will see below,
 this ratio becomes infinitely large as the singular point of collapse is approached.
This means that in the expression: 
\begin{equation}
|\psi _{z}|^{2}=c^{2}[(\partial \psi _{0}/\partial \xi _{\parallel
})^{2}+a^{4}b^{-2}\xi _{\parallel }^{2}\psi _{0}^{2}],\,\,\xi _{\parallel
}=z/a  \label{grad}
\end{equation}%
we can neglect by the first term. Due to the gradient invariance of (\ref{action}), 
which results in the conservation of the quantity $N=\int |\psi_z|^2 d\mathbf{r}$, 
there is a connection between the parameters $a$ and $c$: $c^2(t) = a^5(t)$.
By substituting (\ref{psi}) into (\ref{action}) and performing the subsequent 
integration, we obtain the following expression for the Lagrangian function:
\begin{eqnarray*}
L &=&-I_{0}(a/b)^{2}b_{t}+H(a,b), \\
H(a,b) &=&\widetilde{I_{2}}a^{2}/b^{2}-\widetilde{I_{3}}/(ab^{2})
\end{eqnarray*}%
where 
\begin{equation*}
I_{0}=\frac{1}{2}\int \xi _{\parallel }^{4}\psi _{0}^{2}d\xi _{\parallel }d%
\mathbf{\xi }_{\perp },\,\,\widetilde{I_{2}}=2I_{0},\,\,\,\widetilde{I_{3}}%
=\int \xi _{\parallel }^{4}\psi _{0}^{4}d\xi _{\parallel }d\mathbf{\xi }%
_{\perp },\,\,\mathbf{\xi }_{\perp }=\mathbf{r}_{\perp }/b.
\end{equation*}%
Varying the obtained action 
\begin{equation*}
\widetilde{S}=\int Ldt
\end{equation*}%
relative to $a$ and $b$ we get the following equations of the Euler-Lagrange
equations%
\begin{eqnarray}
I_{0}b^{-2}b_{t} &=&\left( 1/2a\right) \partial H/\partial a,  \label{var-a}
\\
2I_{0}aa_{t} &=&-b^{2}\partial H/\partial b.  \label{var-b}
\end{eqnarray}%
Despite the fact that this analysis is approximate, 
the obtained result is in a good agreement with the exact quasiclassical solution, as shown below.
Equations (\ref{var-a}) and (\ref{var-b}) conserve the system's energy $H=H(a,b)$. 
This conservation law allows us to eliminate $b$ and obtain the dependence of
 $a(t)$ as $b(t) \rightarrow 0$. Asymptotically, $a$ approaches a constant value,
 which is equal to $(\widetilde{I_{3}}/\widetilde{I_{2}})^{1/2}$. In this case, 
$b$ vanishes as $(t_{0}-t)$. Thus, the characteristic feature of this solution is
 the absence of cavity compression along the longitudinal direction.

\section{Quasiclassical description}
As noted in the Introduction, the effective nonlinearity in equation
 (\ref{main}) enables the description of collapse within the 
framework of the quasiclassical approach.
We will seek a solution to equation (\ref{main}) in the following form:

$\psi =A\exp (i\Phi )$ assuming the potential distribution to be
quasiclassical, i.e.%
\begin{equation}
|\Phi _{t}|T\gg 1,  \label{crit-t}
\end{equation}%
\begin{equation}
|\Phi _{z}|L_{z}\gg 1,|\nabla _{\perp }\Phi |L_{\perp }\gg 1,  \label{crit-r}
\end{equation}%
where $T$ is a characteristic time of the variation of the field $A$, 
$L_{z}$ and $L_{\perp }$ are characteristic scales of the amplitude changes
in longitudinal and transverse directions, respectively.

In the leading order relative to these parameters, we get the
Hamilton-Jacobi equation for the eikonal $\Phi $:%
\begin{equation}
\Phi _{t}+\Omega (\nabla \Phi )-n=0,  \label{H-J}
\end{equation}%
where $\Omega (\mathbf{k})=k_{z}^{2}+\mathbf{k}_{\perp }^{2}/k_{z}^{2}$, $%
\mathbf{k=}\nabla \Phi $ are frequency and wave vector, respectively, of
small oscillations, $n=|\psi _{z}|^{2}\approx A^{2}$ $\Phi _{z}^{2}$ is the
field intensity.

In the next order of the perturbation theory, we get the continuity equation
for $n$: 
\begin{equation}
n_{t}+(\nabla \cdot n\mathbf{v)}=0.  \label{cont}
\end{equation}%
Here $\mathbf{v=\partial }\Omega /\partial \mathbf{k}$ ($\mathbf{k=}\nabla
\Phi $) is the group velocity. Equations (\ref{H-J}) and (\ref{cont}) remain
the Hamiltonian structure%
\begin{equation*}
\partial n/\partial t=\delta H/\delta \Phi ,\,\,\partial \Phi /\partial
t=-\delta H/\delta n,
\end{equation*}%
where%
\begin{equation*}
H=\int \left[ n\Omega (\nabla \Phi )-n^{2}/2\right] d\mathbf{r.}
\end{equation*}%
In a hydrodynamical analogy, these equations describe a 
gas of particles with a dependence of particle energy on momentum $\Omega = \Omega(\mathbf{k})$, 
in the presence of negative pressure, which is the main reason for the collapse.

Equations (\ref{H-J}) and (\ref{cont}) admit an entire family of self-similar solutions of the collapsing type: 
\begin{eqnarray}
\Phi (\mathbf{r},t) &=&\int^{t}\frac{\lambda ^{2}dt^{\prime }}{\left(
t_{0}-t^{\prime }\right) ^{\alpha +1}}+\left( t_{0}-t\right) ^{-\alpha
}\varphi _{0}(\mathbf{\xi }),\,\, \lambda ^{2}=const,  \label{phase} \\
n(\mathbf{r},t) &=&\left( t_{0}-t\right) ^{-\alpha -1}f_{0}(\mathbf{\xi }).
\label{intensity}
\end{eqnarray}%
Here self-similar variables $\xi _{\parallel }=z\left( t_{0}-t\right)
^{(\alpha -1)/2}$, $\xi _{\perp }=\mathbf{r}_{\perp }\left( t_{0}-t\right)
^{-1}$, and functions $f_{0}(\mathbf{\xi })$ and $\varphi _{0}(\mathbf{\xi }%
) $ are determined from the corresponding differential equations relative to 
$\mathbf{\xi }$ and positiveness of $f_{0}$ and its decrease at the
infinity, $\mathbf{|\xi |\rightarrow \infty .}$ There is a restriction on
the free parameter $\alpha $ associated with the condition of non-increase
in the number of plasmons in a collapsing cavity with longitudinal, $\propto \left(
t_{0}-t\right) ^{(1-\alpha )/2}$, and transverse, $\propto \left( t_{0}-t\right)$, %
scales, respectively:%
\begin{equation*}
N=\int_{\Omega (t)}n(\mathbf{r}) d {\bf r}\propto \left( t_{0}-t\right) ^{3(1-\alpha )/2},
\end{equation*}%
i.e. $\alpha \leq 1$. The strong collapse regime relates to $\alpha =1$, for
which the longitudinal compression is absent. The weakest and respectively
the most rapid collapse with the index $\alpha =0$ \cite{KrasnoselskikhSotnikov}
corresponds to the boundary of the whole hierarchy of collapsing
solutions where the quasiclassical criteria (\ref{crit-t}) and (\ref{crit-r}%
). This mode describes a self-similar solution of Eq. (\ref{main}) of the
form 
\begin{eqnarray*}
\Phi (\mathbf{r},t) &=&\int^{t}\frac{\lambda ^{2}dt^{\prime }}{%
t_{0}-t^{\prime }}+\varphi _{0}(\mathbf{\xi }),\,\,\lambda ^{2}=const, \\
n(\mathbf{r},t) &=&\left( t_{0}-t\right) ^{-1}f_{0}(\mathbf{\xi }),\,\,\xi
_{\parallel }=z\left( t_{0}-t\right) ^{-1/2},
\,\,\mathbf{\xi }_{\perp }=\mathbf{r%
}_{\perp }\left( t_{0}-t\right) ^{-1}.
\end{eqnarray*}%
Let us show now that the quasiclassical solutions (\ref{phase}) and (\ref%
{intensity}) are unstable with respect to small perturbations. We will seek
for a solution of Eqs. (\ref{H-J}) and (\ref{cont}) in the form%
\begin{eqnarray*}
n &=&\frac{1}{\left( t_{0}-t\right) ^{\alpha +1}}\left( f_{0}(\mathbf{\xi }%
)+\delta f(\mathbf{\xi ,}t)\right) , \\
\Phi (\mathbf{r},t) &=&\int^{t}\frac{\lambda ^{2}dt^{\prime }}{\left(
t_{0}-t^{\prime }\right) ^{\alpha +1}}+\left( t_{0}-t\right) ^{-\alpha
}\left( \varphi _{0}(\mathbf{\xi })+\delta \varphi (\mathbf{\xi ,}t)\right) ,
\end{eqnarray*}%
assuming $\delta f$ and $\delta \varphi $ to be small quantities. We will
suppose now that $\delta f$ and $\delta \varphi $ depend on time like $%
\left( t_{0}-t\right) ^{-\gamma }$. With account of (\ref{phase}) and (\ref%
{intensity}) we arrive at the following spectrum problem:%
\begin{eqnarray}
(\gamma +1+\alpha )\delta f+\frac{\alpha -1}{2}\xi _{\parallel }\frac{%
\partial \delta f}{\partial \xi _{\parallel }}+\mathbf{\xi }_{\perp }\frac{%
\partial \delta f}{\partial \mathbf{\xi }_{\perp }}+\frac{\partial }{%
\partial \xi _{i}}\left( \delta f\frac{\partial \varphi _{0}}{\partial \xi
_{i}}+f_{0}\Omega _{ij}\frac{\partial \delta \varphi }{\partial \xi _{j}}%
\right) &=&0,  \label{lin-1} \\
(\gamma +\alpha )\delta \varphi +\frac{\alpha -1}{2}\xi _{\parallel }\frac{%
\partial \delta \varphi }{\partial \xi _{\parallel }}+\mathbf{\xi }_{\perp }%
\frac{\partial \delta \varphi }{\partial \mathbf{\xi }_{\perp }}+\frac{%
\partial \Omega }{\partial k_{i}}\frac{\partial \delta \varphi }{\partial
\xi _{i}}-\delta f &=&0,  \label{lin-2}
\end{eqnarray}%
where $\Omega _{ij}=\partial ^{2}\Omega /\partial (k_{i}k_{j})$. If this
system has a solution with $\gamma >0$ then distribution (\ref{phase}) and (%
\ref{intensity}) will be unstable, and stable in the opposite case.
The simplest way to determine the spectrum $\gamma$ is by considering shortwave perturbations, 
in which the dependencies of $f_0$ and $\varphi_0$ on the coordinates $\mathbf{\xi}$ can be neglected. 
In this case, the perturbations $\delta f$ and $\delta \varphi$ depend on $\mathbf{\xi}$ exponentially, 
$\sim \exp (i\mathbf{p\xi})$ ($\mathbf{|p|\gg }1$). Thus, the "growth rate" $\gamma$ can be expressed with sufficient accuracy as:
\begin{equation*}
\gamma = \left\vert f_{0}\Omega _{ij}p_{i}p_{j}\right\vert ^{1/2}.
\end{equation*}
This indicates instability, which becomes stronger as the perturbation wavelength increases.
 For smooth initial conditions, the instability will become significant at a later time than 
for more oscillating initial conditions. As a matter of fact, the obtained instability is essentially 
the modulation instability of the given distribution. It should be stressed that the same method 
was used to obtain the expression for the growth rate of instability in quasiclassical solutions 
of the NLSE \cite{ZakharovKuznetsov1986}. It is noteworthy that all quasiclassical solutions 
are unstable, except possibly for the solution (\ref{phase}) and (\ref{intensity}) with $\alpha = 0$. 
For distributions with $\alpha = 0$, the quasiclassical conditions
 (\ref{crit-t}) and (\ref{crit-r}) are violated, and the instability apparently disappears.

 \section{Anisotropic "black holes"}
The obtained results suggest the following scenario for the development of collapse.
 The instability of the quasiclassical solutions develops on the background of weak 
singularities in the compressing wave packet. The energy captured in each 
singularity is formally equal to zero and depends on the nonlinear damping, 
which is associated either with the transition of collapse to the hydrodynamic 
stage or with Landau damping. Previous studies on the NLS equation, 
such as \cite{VlasovPiskunovaTalanov}, \cite{Malkin}, and \cite{ZakharovKosmatovShvets}, 
have shown that the singularity formed during a weak collapse does not disappear after 
its completion but continues to exist, absorbing finite energy into itself. 
The situation is similar to the solution of equation (\ref{main}). 
The singularity, in the form of a "burning point," which appears during a weak collapse, 
can continue to exist, quasi-stationarily absorbing wave energy from the packet. 
In contrast to the NLS equation, the "black hole" formed within 
the framework of (\ref{main}) has an anisotropic structure. This situation corresponds 
to solutions of Eqs. (\ref{H-J}) and (\ref{cont}), which are singular at a point in the form of an anisotropic funnel.

\begin{equation}
\Phi (r_{\perp },z)=z^{-1}\chi (\eta ),\,\,n(r_{\perp },z)=z^{-4}g(\eta ).
\label{hole}
\end{equation}%
Here $\eta =r_{\perp }/z^{2}$, and function $g(\eta )$, as it can be easily
shown, is satisfied the ordinary differential equation%
\begin{equation}
gg^{\prime }+3\eta (g+3\eta g^{\prime })^{4}=0.  \label{black}
\end{equation}%
Solution of this nonlinear equation of the first order  depends on one
constant. As such a constant, one can choose a quantity%
\begin{equation*}
P=\int_{0}^{\infty }(1-3\eta g^{\prime }g^{-1})g^{\prime }d\eta
\end{equation*}
that has the meaning of an energy flux into a singularity.

Numerical modeling confirmed the existence of a monotonically
decreasing solution of Eq. (\ref{black}). For \ $\eta \rightarrow \infty $,
its asymptotics has the form: $g\sim \eta ^{-1/2}$. It is easy to see that
the quasiclassical criteria of solution (\ref{hole}) improve
while approaching the singularity $|\mathbf{r}|=0$. 

Note that stationary solutions of Eq. (\ref{main}) singular at
the origin have, in comparison with (\ref{hole}), a weaker singularity%
\begin{equation*}
|\psi _{z}|^{2}=z^{-2}h(r_{\perp }/z^{2}).
\end{equation*}%
Therefore, it is expected that collapse leads to mode (%
\ref{hole}).

In conclusion, the authors thank V.E. Zakharov for useful discussions and
the opportunity to get acquainted with paper \cite{ZakharovKosmatovShvets}
before its publication, as well as V.K. Mezentsev for performing numerical
verifications.

\end{document}